\documentclass[prd,preprint]{revtex4}
\usepackage{amsmath}
\usepackage{amssymb}
\usepackage{graphicx}
\usepackage{psfig}
\newcommand{\be}{\begin{eqnarray}}
\newcommand{\ben}{\begin{eqnarray}\nonumber}
\newcommand{\ee}{\end{eqnarray}}

\begin{document}

\title{Cooling of Neutron Stars with Color Superconducting\\ Quark Cores}

\author{Hovik Grigorian$^{1,2}$, David Blaschke$^{3,4}$, and
Dmitri Voskresensky$^{5,6}$}

\affiliation{$^1$Fachbereich Physik, Universit\"at Rostock, 
D-18051 Rostock, Germany\\
$^2$Department of Physics, Yerevan State University, Alex
        Manoogian Str. 1, 375025 Yerevan, Armenia\\
$^3$Fakult\"at f\"ur Physik, Universit\"at
Bielefeld, D-33615 Bielefeld, Germany \\
$^4$Bogoliubov Laboratory of Theoretical Physics,
Joint Institute for Nuclear Research, 141980, Dubna, Russia\\
$^5$Theory Division, GSI mbH, D--64291 Darmstadt, Germany\\
$^6$Moscow Institute for Physics and Engineering, 115409 Moscow, 
Russia}

%%\thanks[VI]{Supported by the Virtual Institute of the Helmholtz Association 
%%under grant No. VH-VI-041}

\begin{abstract} 
We show that within a recently developed nonlocal chiral quark model
the critical density for a phase transition to color superconducting
quark matter under neutron star conditions can be low enough for these
phases to occur in compact star configurations with masses below 
$1.3~M_\odot$.
We study the cooling of these objects in isolation for different values
of the gravitational mass.
%% and thus different composition and structure of the interior.
Our equation of state (EoS) allows for 2SC quark matter with a large quark
gap $\sim 100~$MeV for $u$ and $d$ quarks of two
colors that coexists  with normal quark matter within a mixed phase
%%and a pure normal quark phase
%%within 
in the hybrid star interior.  We argue that, if  the phases
with unpaired quarks were allowed, the corresponding hybrid stars
would cool too fast. If they occured for  $M<1.3~M_\odot$, as it follows from
our EoS, one could not  appropriately describe the neutron star cooling data
existing by today. We discuss a "2SC+X"  phase, as a possibility to have all 
quarks paired in two-flavor quark matter under neutron star constraints, where 
the X-gap is of the order of 10 keV - 1 MeV. 
Density independent gaps do not allow to fit the cooling data. 
Only the presence of
an X-gap that decreases with increase of the density
%%density dependence of this weak residual pairing channel
could allow to appropriately  fit the data in a similar compact 
star mass interval to that following from a purely hadronic model. 
This scenario is suggested as an alternative explanation of the cooling data 
in the framework of a hybrid star model.
\end{abstract}

%Neutron stars \sep Neutrino cooling  \sep Color superconductivity

\maketitle

\section{Introduction}

%%The high-density phases of QCD at low temperatures may exist in
%%the interiors of hybrid stars affecting their cooling, rotation and
%%magnetic field evolution, cf. \cite{bkv,ppls,bgv,bss,IB}.

%%In the recent paper \cite{bgv2004}
%%, hereafter BGV, 
Recently, we have reinvestigated
the cooling of neutron stars  within a purely hadronic model \cite{bgv2004}, 
i.e., suppressing the possibility of quark cores in neutron star interiors.
We have demonstrated that the neutron star cooling data available by today 
can be well explained within the {\em "Nuclear medium cooling scenario"}, 
cf. \cite{SVSWW,V01}, i.e., if one includes medium effects into consideration.
%% and takes into account a suppression of  the $3P_2$ neutron gap.
This scenario puts a tight constraint to the density dependence
of the asymmetry energy in the nuclear equation of state (EoS). 
The  latter dependence is an important issue for
the analysis of heavy ion collisions especially within the new CBM 
(compressed baryon matter) program to be realized at the future accelerator
facility FAIR at GSI Darmstadt. 
The density dependence of the asymmetry energy
determines the proton fraction in neutron star matter and thus governs the 
onset of the very efficient direct Urca (DU) process. 
This process, once occurring, 
would lead to a very fast cooling of neutron stars.
%% in disagreement with available cooling data. 
Thereby, if neutron stars with $M<M_{\rm crit}^{\rm DU}$ cooled  slowly, the
stars with the mass $M$ only slightly above $M_{\rm crit}^{\rm DU}$ would be
already characterized by a very rapid cooling, yielding surface 
temperatures significantly below those given by the $X$-ray data points. 
Unfortunately, the mass spectrum of cooling neutron stars is not known.
One can not rely on neutron star
mass measurements in binary radio pulsars (BRPs) with the averaged neutron star
mass value $M_{\rm BRP}=1.35\pm 0.04~M_{\odot}$ \cite{thorsett,L04} because 
these objects come from a twice selected population: probably not all neutron 
stars go through the active radio-pulsar stage \cite{gotthelf}, and the 
properties of neutron stars in binaries may be different from those of 
isolated objects, see e.g. \cite{pod+}.
However, it seems not a fair assumption that most of cooling neutron stars 
should  be essentially lighter than 
%%a typical neutron star in a binary system,
%%for which masses around 
the lower limit for BRP masses $M_{\rm BRP, min}=1.31~M_\odot$.
Therefore, the threshold for the DU process should better occur at high enough
densities such that  typical neutron stars are not affected by it.
This assumption about the mass distribution can be developed into a more
quantitative test of cooling scenarios when these are combined with 
population synthesis models. 
The latter allow to obtain Log N -- Log S distributions
for nearby coolers which can be tested with data from the ROSAT catalogue
\cite{pgtb2004}.
In addition the microscopically based Argonne model for the EoS, that is 
supposed to be the most realistic one \cite{APR98}, yields a critical mass 
for the DU process $\sim 2~M_{\odot}$.
At these high star masses the central baryon density exceeds already the 
5-fold of the nuclear saturation density such that exotic states of matter
as, e.g., hyperonic matter or quark matter are more appropriate.
Ref. \cite{Baldo:2003vx} argued that the presence of quark matter in massive 
compact star cores is a most reliable hypothesis  
since within their model hyperons appear at sufficiently low densities and the
equation of state becomes too soft to carry neutron star masses in the range 
of $M_{\rm BRP}$ \cite{footnote}. 
  
For quark matter in neutron stars, the
%%re cannot be a large enough asymmetry to prevent 
interaction ($\alpha_s \neq 0$) permits the quark DU process. The  works
\cite{bkv,ppls,bgv,bgv2004q} have shown that this result does
not exclude the possibility that neutron stars might possess large
quark matter cores. 
%%that extend up to more than half of the star radius. 
The
suppression of the emissivity of the DU process may
arise due to the occurence of diquark pairing and thus color superconductivity
in low-temperature quark matter.
%%, cf.\cite{bkv,ppls,bgv,bgv2004q}.
To demonstrate this possibility one has used the simplifying assumption 
that all quarks are gapped with the same gap which is not valid for 
superconducting phases at low densities. 
In the present paper we abandon that restriction and suggest a more subtle
pairing pattern that is constrained by comparison with neutron star cooling
phenomenology. 

%Such a hybrid structure gives room for a whole variety of additional
%scenarios of compact star cooling which fall into two classes:
%either nuclear and quark matter phases have similar cooling behavior
%(homgeneous cooling) or the faster cooling of the one phase is compensated
%by the slower cooling of the other (inhomogeneous cooling).
%In the present contribution we will report on our results within the former,
%homgeneous cooling scenario of hybrid stars and what implications the
%comparison with present-day cooling data may provide for the EoS and
%transport properties of quark matter.

\section{Color superconductivity}

The quark-quark interaction in the color anti-triplet channel is
attractive driving the pairing with a large zero-temperature
pairing gap $\Delta\sim 100$~MeV for the quark chemical
potential $\mu_q \sim (300\div 500)$~MeV, cf.
%%, cf. \cite{bl}. The problem has
%%been re-investigated in a series of papers following Refs.
\cite{arw98,r+98}, for a review see \cite{RW} and references therein.
The attraction comes either from the one-gluon exchange, or from a
non-perturbative 4-point interaction motivated by instantons
\cite{dfl}, or from non-perturbative gluon propagators \cite{br}.
Various phases are possible. The so called 2-flavor color
superconductivity (2SC) phase allows for unpaired quarks of one
color, say blue. There may also exist a color-flavor locked
(CFL) phase \cite{arw99} for not too large values of the dymanical
strange quark mass or in other words for very large values of the baryon
chemical potential \cite{abr99}, where the color superconductivity
is complete in the sense that the diquark condensation
produces a gap for quarks of all three colors and flavors. The
value of the gap is of the same order of magnitude as that in the
two-flavor case. There exist other attractive quark pairing
channels. However the attraction is rather weak. One of these channels allows
for a color spin locking (CSL),
where all  quarks are paired in a spin-1 channel.
%%for quarks that can't participate in
%%2SC and CFL pairing.
The weak pairing channels are characterized by gaps
typically in the interval $\sim 10~$ keV $\div 1~$MeV.
%Other phases allow for unpaired quarks within so called gapless CSC, see
%discussion of different attractive interaction channels in paper \cite{ABCC}.
%%Below we will discuss the possibility of the pairing of previously unpaired
%%quarks of the normal quark phase and unpaired blue quarks
%%of the 2SC phase within that we had within the mixed phase
%%due to a possible weak pairing mechanism.
%%We will not specify the microscopic origin of this mechanism
%%\%% residual pairing in the present work. 
%%and therefore we call the corresponding phase 
%%2SC+X with a large 2SC gap $\Delta \sim 100$ MeV and a small  X-gap
%%$\Delta_X\sim 10$ keV $\div 1$ MeV. Instead we will confront this possibility
%%with the cooling data.

\section{Hybrid stars}

In describing the hadronic part of the hybrid star, as in  \cite{bgv2004}, we  
exploit the Argonne $V18+\delta v+UIX^*$ model of the EoS given in 
\cite{APR98}, which is based on the most recent models for the nucleon-nucleon 
interaction with the inclusion of a parameterized three-body force and 
relativistic boost corrections.
Actually we continue to adopt an analytic parameterization of this model by 
Heiselberg and Hjorth-Jensen \cite{HJ99}, hereafter HHJ,
%%. The fit is done for $n<4~n_0$.
%%The HHJ EoS uses a compressional part with the incompressibility 
%%$K\simeq 240$~MeV,
%%and a symmetry energy fitted to the data around nuclear saturation density 
that smoothly incorporates causality at high densities. 
%%The density dependence
%%of the symmetry energy is very important since it determines the
%%value of the threshold density for the DU process ($n_c^{\rm DU}$).
The HHJ EoS  fits the symmetry energy  to the original
Argonne $V18+\delta v +UIX^*$ model in the mentioned density interval yielding 
the threshold density for the DU process 
$n_c^{\rm DU}\simeq~5.19~n_0$ ($M_c^{\rm DU}\simeq 1.839~M_{\odot}$).

The 2SC phase occurs at lower baryon densities than the CFL phase,
see \cite{SRP,NBO}. For applications to compact
stars the omission of the strange quark flavor is justified by the
fact that chemical potentials in central parts of the stars do
barely reach the threshold value at which the mass gap for strange
quarks breaks down and they appear in the system \cite{GBKG}.

We will employ the EoS of a nonlocal chiral quark model developed in
\cite{BFGO} for the case of neutron star constraints with a 2SC phase. 
It has been shown in that work that the Gaussian formfactor ansatz leads 
to an early onset of the deconfinement transition and such a model is 
therefore suitable to discuss hybrid stars with large quark matter cores
\cite{GBA}.

In some density interval above the onset of the first order phase transition,
there may appear a mixed phase region, see \cite{G92}. Ref.
\cite{G92} disregarded finite size effects, such as surface
tension and charge screening. Refs \cite{VYT02} on the example of
the hadron-quark mixed phase have demonstrated that finite size
effects might play a crucial role substantially narrowing the
region of the mixed phase or even forbidding its appearance
in a hybrid star configuration.
Therefore we omit the possibility of the hadron-quark mixed phase in our
model assuming that the quark phase arises by the Maxwell construction. In
this case we found a tiny  
density jump on the boundary from $n_c^{\rm hadr}\simeq 0.44~fm^{-3}$ to
{\bf{$n_c^{\rm quark}\simeq 0.46~fm^{-3}$.}}

A large difference between chemical potentials of $u$ and $d$
quarks forbids the pure 2SC phase, cf. \cite{BFGO}. 
The CFL phase is still not permitted at such
densities. 
Ignoring weak pairing patterns  there could be two possibilities: 
either the quark matter is in the normal phase, or there appears a 
region of the 2SC -- normal quark mixed phase. 
Disregarding finite size effects, as in \cite{G92}, 
Refs. \cite{NBO,GBA} have argued for a broad region of the
2SC -- normal quark mixed phase 
instead of a pure 2SC phase. 
In our model the mixed phase continues up to the central density $n_{centr}$.
Thus we have the mixed phase in the density interval:
$n_c^{\rm hadr}< n_c^{(mix,1)}< n_c^{\rm quark}<n<n_c^{(mix,2)}=n_{centr}$.
In the given case the arguments of \cite{VYT02} might be partially relaxed 
since the surface
tension for the 2SC -- normal quark boundary should be significantly smaller
than for the quark -- hadron boundary. 
Indeed, as has been indicated in Ref. \cite{bgv2004q}, the
surface tension is proportional to the difference of the energies
of two phases, being proportional to  $(\Delta/\mu_q )^2 \ll 1$ 
in the given case. 
Recently, Ref. \cite{RR} estimated the surface tension to be typically 
smaller than $5$~MeV$/$fm$^2$ for densities of our interest. 
For the choice of the Gaussian formfactor the mixed phase appears for
sufficiently low density 
$n> n_c^{(mix,1)}\simeq n_c^{\rm hadr}=0.44 ~{\rm fm}^{-3}$ 
($M>M_c^{\rm quark}=1.214~M_\odot$). 
The presence or absence of the 2SC - normal quark mixed phase
instead of only one of those phases
is not so important for the hybrid star cooling problem since the latter is
governed by processes involving either normal excitations or excitations 
with the smallest gap.

\begin{figure}[ht]
%\vspace{-0.5cm}
\centerline{
\psfig{figure=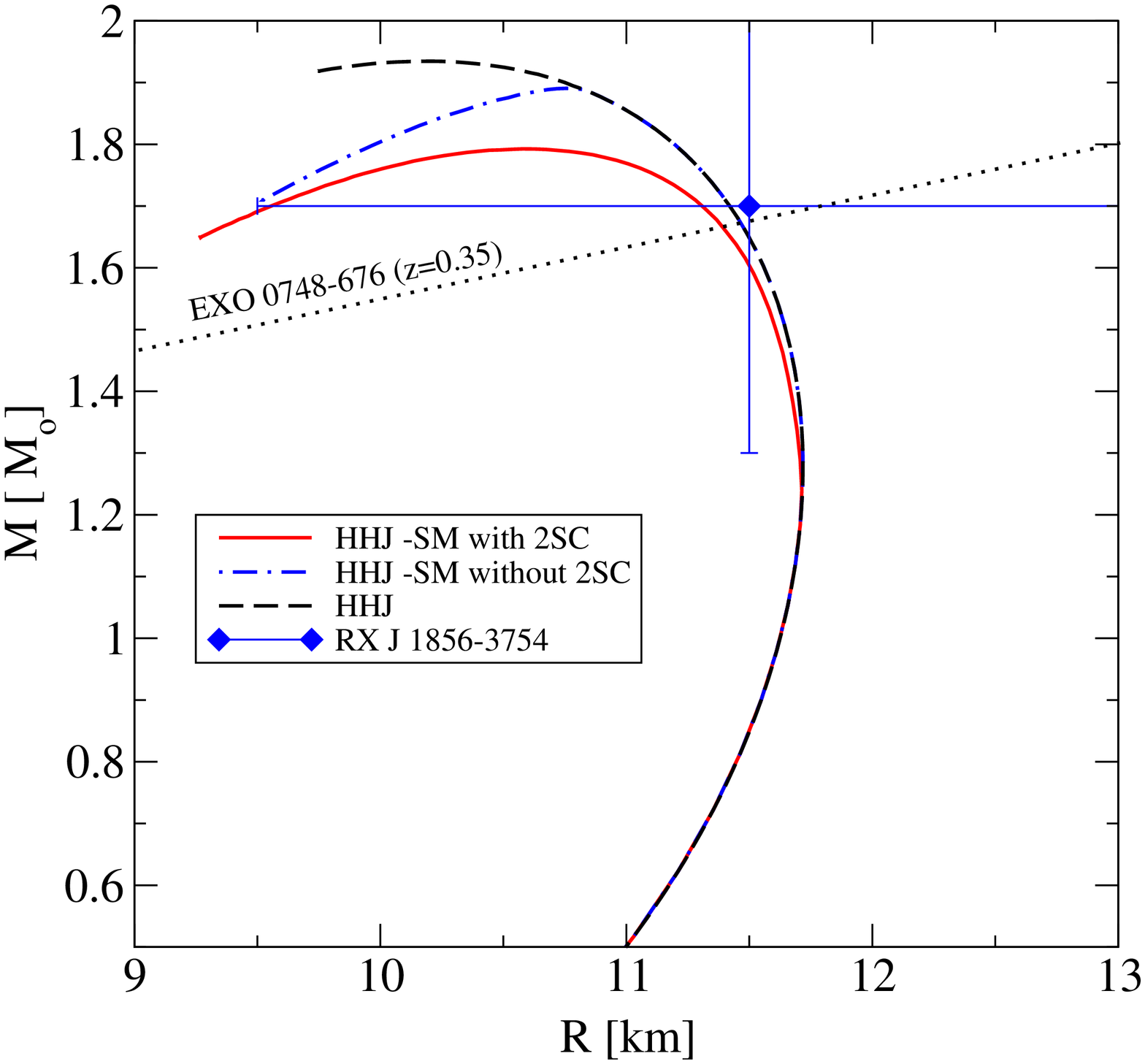,height=0.85\textwidth,angle=-90}}
\caption{Mass - radius 
relations for compact star configurations with different EoS:
purely hadronic star with HHJ EoS (dashed line),
stable hybrid stars with HHJ - Gaussian nonlocal chiral quark separable
model (SM) 
with 2SC phase (solid line) and
with HHJ - SM, without 2SC phase (dash-dotted line). 
Data for two sources are also indicated, see text.
\label{fig:stab}}
\end{figure}

In Fig. \ref{fig:stab} we present the mass-radius relation for hybrid stars
with HHJ EoS vs. Gaussian nonlocal chiral quark separable model (SM) EoS. 
Configurations with the 2SC - normal quark mixed phase, ``HHJ-SM with 2SC'' 
(HHJ hadronic EoS together with Gaussian nonlocal chiral quark model SM 
permitting 2SC - normal quark mixed phase) given by the solid line, 
are stable, whereas without 2SC large gap color superconducting
matter (``HHJ-SM without 2SC'') no stable hybrid star configuration is 
possible. In the case ``HHJ-SM with 2SC'' the maximum neutron star mass proves 
to be $1.793~M_{\odot}$. 
For an illustration of the constraints on the mas-radius relation 
which can be derived from compact star observations we show in Fig. 
\ref{fig:stab} the compactness limits from the thermal emission of 
the isolated neutron star RX J1856.5-3754 as given in \cite{PLSP} 
and from the redshifted absorption lines in the X-ray burst spectra 
of EXO 0748-676 given in \cite{CPM}. 
These are, however, rather weak  constraints.

Additionally, within the ``HHJ-SM with 2SC'' mixed phase
we will allow for the possibility of a weak pairing channel
for all the  quarks which were unpaired, 
with typical gaps  $\Delta_X \sim 10$~keV $\div 1$~MeV, as in the case 
of the CSL pairing channel, see \cite{Schafer,Schmitt}.
Since we don't know yet the exact pairing pattern for this case,
we call this hypothetical phase ``2SC+X''.
%Since typical sizes
%of the droplets and of the Wigner-Setz cells are rather small
%($\sim 1\div 10$~fm) compared to the typical size on which quarks
%undergo a weak pairing ($\sim 1/\Delta$),  red quarks from 2SC
%region may pair with  blue and green quarks of the normal phase.
In such a way all the quarks get paired, some strongly in the 
2SC channel and some weakly in the X channel.

\section{Cooling}
For the calculation of the cooling of the hadron part of the
hybrid star we use the same model as in \cite{bgv2004}.
The main processes are the medium modified Urca (MMU) and the pair
breaking and formation (PBF) processes. The HHJ EoS is adopted.
Below we choose two cooling  models used in \cite{bgv2004} distinguished
by different sets of the nucleon gaps. 
The first choice (further model I) is shown in  Fig. 2 
(Fig. 12 of Ref. \cite{bgv2004}). 
We use a fit for the relation between surface and crust
temperatures (called ``our fit'' in \cite{bgv2004}). 
We also made calculations for the other two crust models used in 
Ref. \cite{bgv2004}. Although the results are affected by
the choice of the crust model, our conclusions remain untouched.
The possibilities of pion condensation and of other so called exotic processes 
are suppressed since in the model \cite{bgv2004} these processes may occur only
for neutron star masses exceeding $M_c^{\rm quark}= 1.214~M_{\odot}$. 
The DU process is irrelevant in this model up
to very large neutron star mass $M>1.839~M_{\odot}$. 
The $1S_0$ neutron and proton gaps are taken the same as those shown by thick 
lines in Fig. 5 of  Ref. \cite{bgv2004}. 
We pay particular attention to the fact that the 
$3P_2$ neutron gap is additionally suppressed by the factor $0.1$ compared to 
that shown in Fig. 5 of \cite{bgv2004}. 
The latter suppresion  is motivated by  the result
of the recent work \cite{SF} and is required to fit the cooling data. 
%%Below we also use another choice for the gaps (model II),  as shown by
%%dash lines in Fig. 5 of \cite{bgv2004}. 
%%We will see that although the results are essentially model
%%dependent it does not change our qualitative conclusions. 

For the calculation of the cooling of the quark core in the hybrid
star we use the model \cite{bgv}. We incorporate the most
efficient processes: the quark direct Urca (QDU) processes on
unpaired quarks, the quark modified Urca (QMU), the quark
bremsstrahlung (QB), the electron bremsstrahlung (EB), and the massive
gluon-photon decay (see \cite{bkv}). Following
\cite{JP02} we include the emissivity of the quark pair formation
and breaking (QPFB) processes. The specific heat incorporates the
quark contribution, the electron contribution and the massles and massive 
gluon-photon contributions. 
The heat conductivity contains quark, electron and gluon terms.

\begin{figure}[ht]
%\vspace{-0.5cm}
\centerline{
\psfig{figure=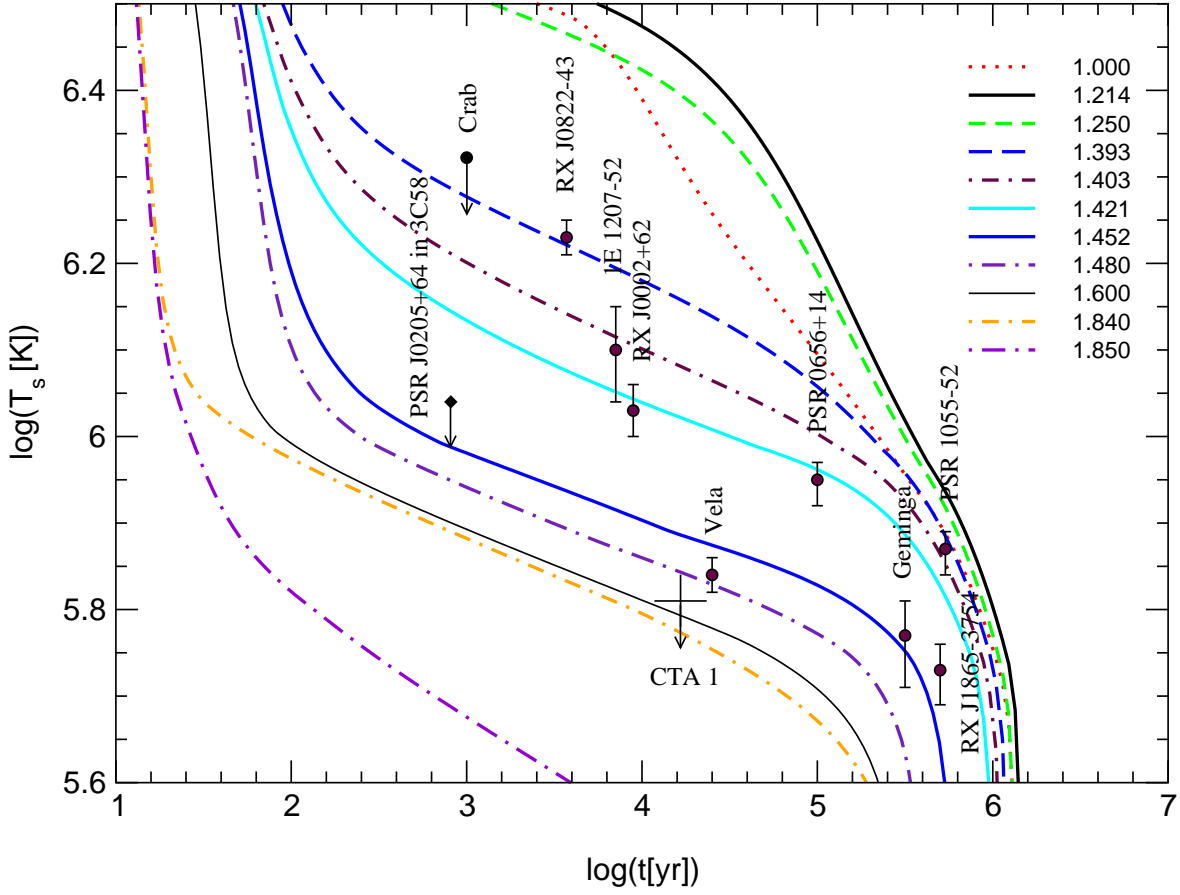,height=0.9\textwidth,angle=-90}}
\caption{Model I. Cooling curves according to the nuclear medium cooling 
scenario, see Fig. 12 of \cite{bgv2004}.
The labels correspond to the gravitational masses of the configurations 
in units of the solar mass.}
\label{fig:cool-h}
\end{figure}

We are basing on the picture presented in Fig. \ref{fig:cool-h} and add the
contribution of the quark core. 
For the Gaussian formfactor the quark core occurs already for
$M>1.214~M_\odot$ according to the model \cite{BFGO}, see Fig. \ref{fig:stab}.
Most of the relevant neutron star configurations (see Fig. \ref{fig:cool-h})
are then  affected by the presence of the quark core.
First we check the possibility of the 2SC+ normal quark phases.

\begin{figure}[ht]
%\vspace{-0.5cm}
\centerline{
\psfig{figure=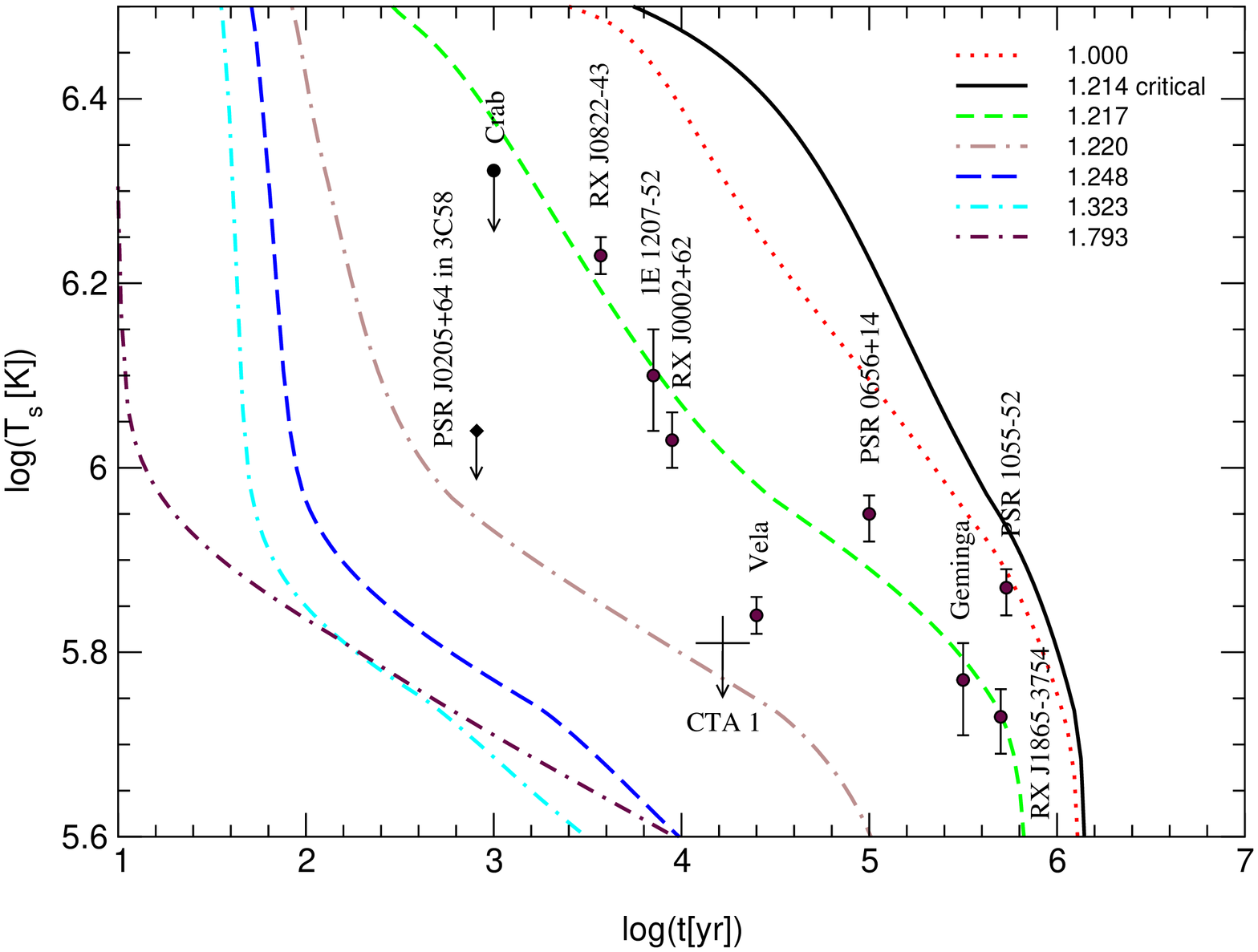,height=0.9\textwidth,angle=-90}}
\caption{Model I. Cooling curves for hybrid star configurations with Gaussian 
quark matter core in the 2SC phase.
The labels correspond to the gravitational masses of the configurations 
in units of the solar mass.}
\label{fig:cool-2sc}
\end{figure}

Fig. \ref{fig:cool-2sc} shows the cooling curves calculated with the given
Gaussian ansatz.  The variation of the gaps for the strong pairing of
quarks within the 2SC phase and the gluon-photon mass in the interval 
$\Delta, m_{g-\gamma}\sim 20\div 200~$MeV only slightly affects the results. 
The main cooling process is the QDU process on normal quarks. We see that
the presence of normal quarks entails too fast cooling. The data
could be explained only if all the masses lie in a very narrow
interval ($1.21<M/M_\odot<1.22$ in our case). In case of the other two  crust
models the resulting picture is similar.
%The value $M\simeq ???$ depends on the model for the form-factor 
%and can be shifted. However 
The existence of only a very narrow mass interval in which the
data can be fitted seems us unrealistic as by itself as from the
point of view of the observation of the neutron stars in binary systems with 
different masses, e.g.,
$M_{\rm B1913+16}\simeq 1.4408 \pm 0.0003~M_{\odot}$ and 
$M_{\rm J0737-3039B}\simeq 1.250 \pm 0.005~M_{\odot}$, 
cf. \cite{L04}. {\em{Thus the data can't be satisfactorily explained.}}

\begin{figure}[ht]
%\vspace{-0.5cm}
\centerline{
\psfig{figure=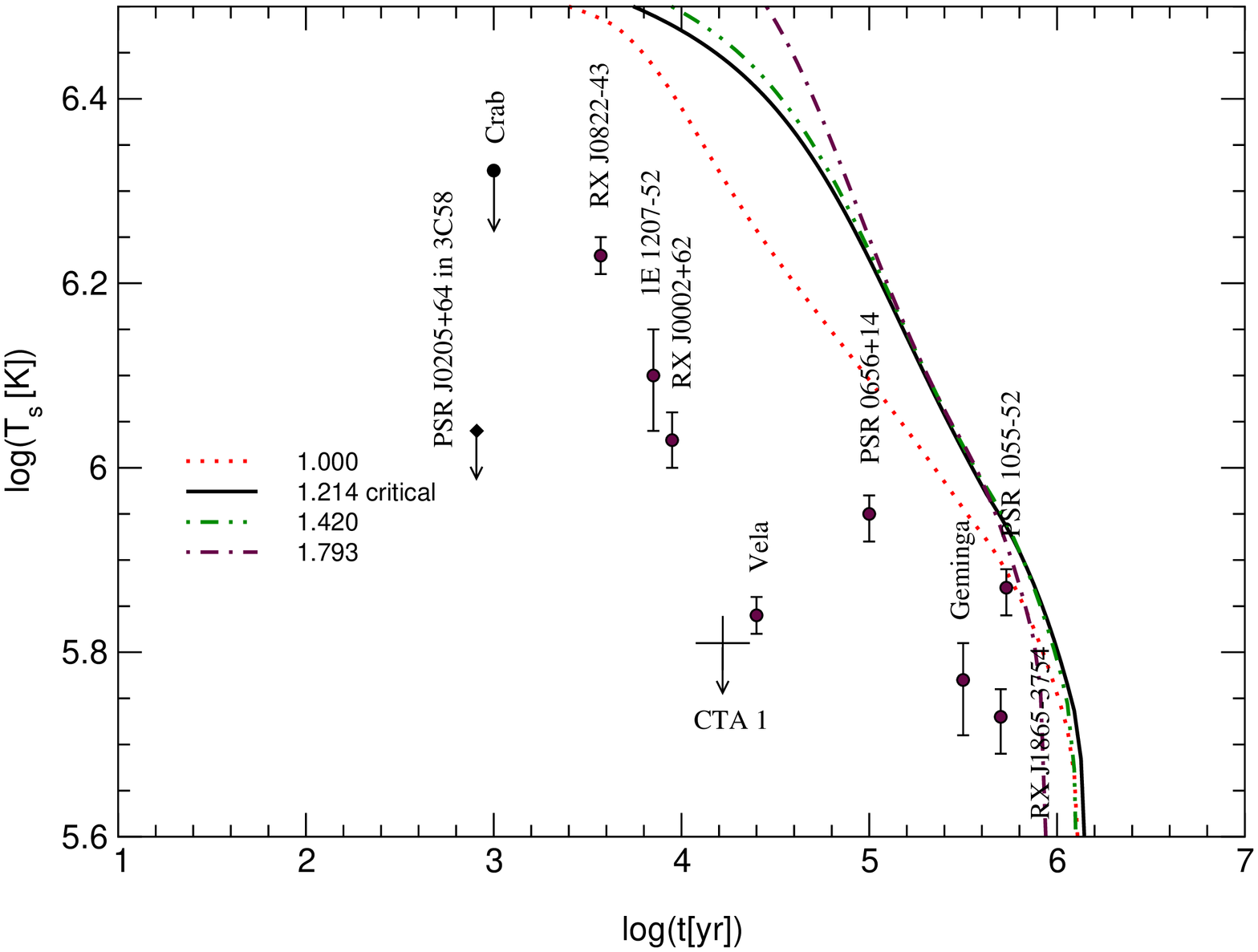,height=0.9\textwidth,angle=-90}}
\caption{Model I. Cooling curves for hybrid star configurations with Gaussian 
quark matter core in the 2SC+X phase. The weak pairing gap is $\Delta_X \simeq 1$~MeV.
The labels correspond to the gravitational masses of the configurations 
in units of the solar mass.}
\label{fig:cool-csl1}
\end{figure}

In Fig. \ref{fig:cool-csl1} we permit the weak pairing pattern for all quarks 
which before in Fig. 3 were assumed to be unpaired. 
We use $\Delta_X \simeq 1~$MeV for the corresponding quark gap. 
Fig. \ref{fig:cool-csl1} demonstrates too slow cooling. 
Even slow cooling data are not explained. 
The same statement holds for all three crust models. Thus the gaps for
formerly unpaired quarks should be still smaller in order to obtain a 
satisfactory description of the cooling data.

\begin{figure}[ht]
%\vspace{-0.5cm}
\centerline{
\psfig{figure=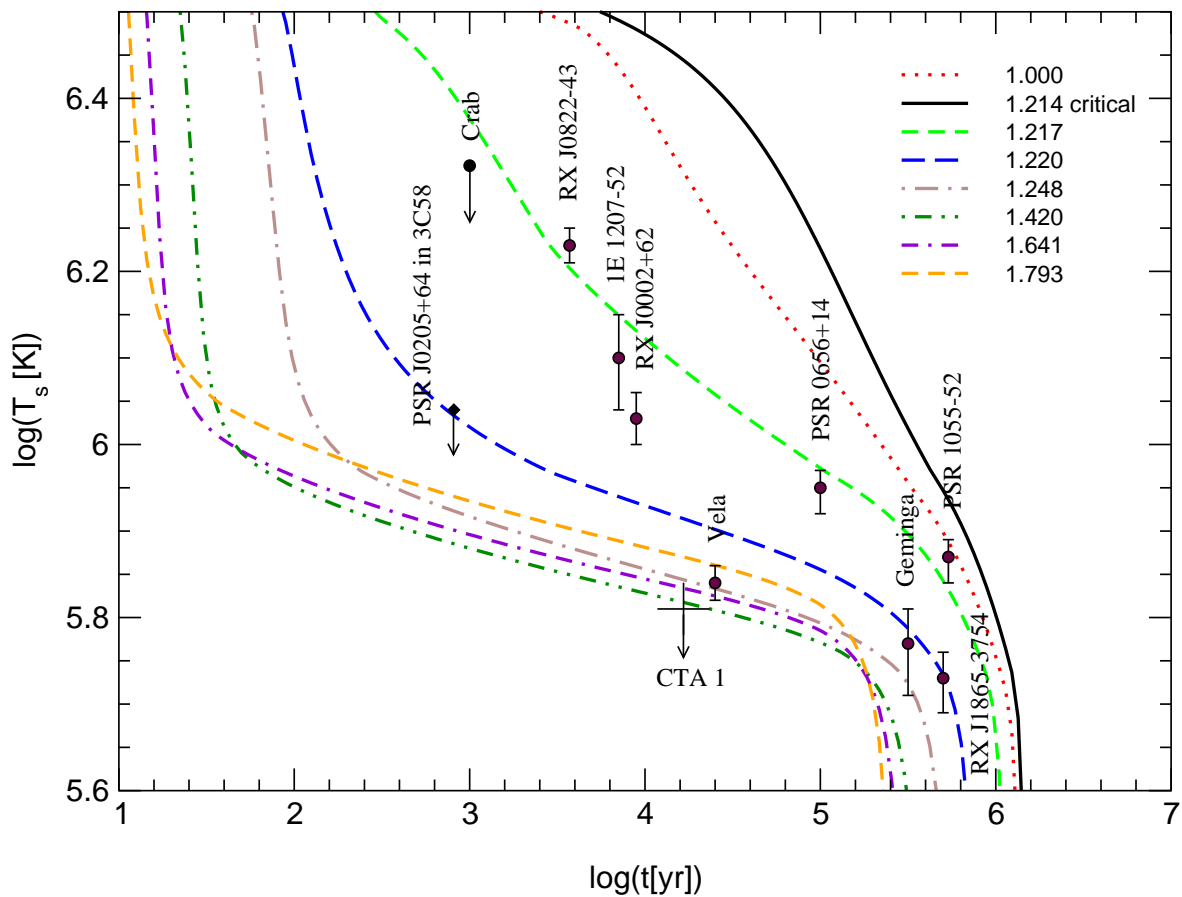,height=0.9\textwidth,angle=-90}}
\caption{Same as Fig. \ref{fig:cool-csl1} but 
%using the Tsuruta law.
with a weak pairing gap of 30 keV.}
\label{fig:cool-csl003}
\end{figure}

In Fig. \ref{fig:cool-csl003} we again allow for the weak pairing pattern
of those quarks which were assumed to be unpaired in
Fig. \ref{fig:cool-2sc}, but now we use
$\Delta_X = 30~$ keV for the corresponding residual quark pairing gap (X). 
%%The cooling data can be  fitted but have a very fragile dependence on the 
%%gravitational mass of the configuration. 
We see that all data points, except the Vela, CTA 1 and Geminga, correspond to
hybrid stars with masses in the narrow interval $M=1.21\div 1.22 ~M_\odot$, 
very similar to that in  Fig. \ref{fig:cool-2sc}.
This also seems to be unrealistic with the same argument as for the presence of
unpaired quarks, see Fig. 3 above. 
We have obtained a similar picture for  the other two  crust models.
We have varied the constant $\Delta$ in wide limits but
the qualitative picture does not change. 

\begin{figure}[ht]
%\vspace{-0.5cm}
\centerline{
\psfig{figure=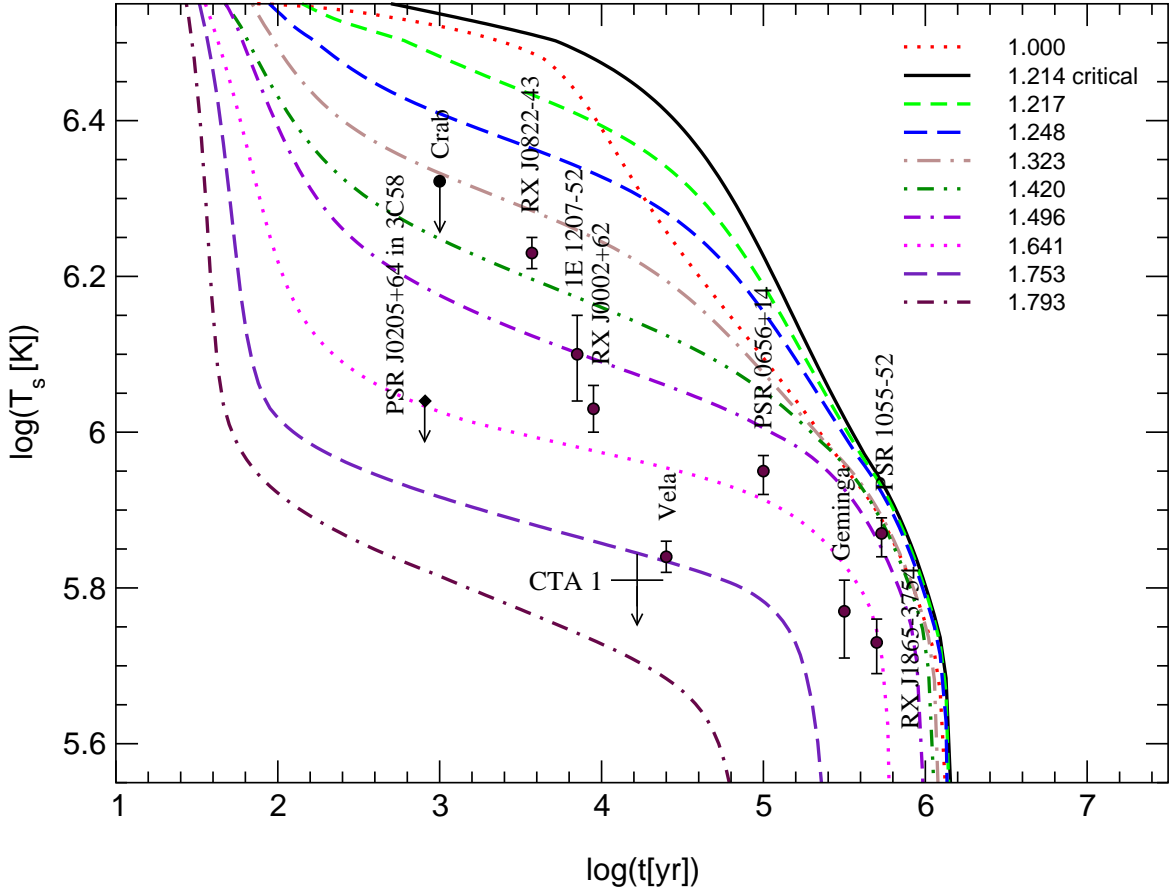,height=0.9\textwidth,angle=-90}}
\caption{Same as Fig. \ref{fig:cool-csl1} but 
%using the Tsuruta law.
with a density dependent pairing gap according to Eq. (\ref{gap}).}
\label{fig:cool-csl-x}
\end{figure}

Therefore we would like to explore whether a
density-dependent X-gap could allow a description of the cooling data 
within a larger interval of compact star masses. 
%%For such a density-dependent X-gap function w
We don't choose the X-gap as a rising function of the chemical
potential
since the feature of such a model will not spread the accessible mass
interval satisfying the cooling data constraint relative to the
density independent X-gap case. We employ the ansatz
\begin{equation}
\label{gap}
\Delta_X(\mu)=\Delta_c~\exp[-\alpha(\mu-\mu_c)/\mu_c]~,
\end{equation}
where the parameters are chosen such that at the critical quark chemical 
potential $\mu_c=330$ MeV for the onset of the deconfinement phase 
transition the X-gap has its maximal value of $\Delta_c=1.0$ MeV 
and at the highest attainable chemical potential $\mu_{\rm max}=507$ MeV, 
i.e. in the center of the maximum mass hybrid star configuration it 
falls to a value of the order of $10$ keV.
We choose the value $\alpha=10$ for which 
$\Delta_X(\mu_{\rm max})=4.6$ keV.
In Fig. \ref{fig:cool-csl-x} we show the resulting cooling curves for 
the ansatz (\ref{gap}). 
We observe that the mass interval for compact stars which obey the 
cooling data constraint ranges now from $M=1.32~M_\odot$ for slow
coolers up to $M=1.75~M_\odot$ for fast coolers such as Vela, cf. with that we
have found with the purely hadronic model \cite{bgv2004} with different 
parameter choices.
Note that according to a recently suggested independent test of cooling 
models \cite{pgtb2004} by comparing results of a corresponding population 
synthesis model
with the Log N - Log S  distribution of nearby isolated X-ray sources
the cooling model I  did not pass the test. 
Thereby it would be interesting to see whether our quark model within the gap 
ansatz (\ref{gap}) may pass  the  Log N - Log S test.

\begin{figure}[ht]
\centerline{
\psfig{figure=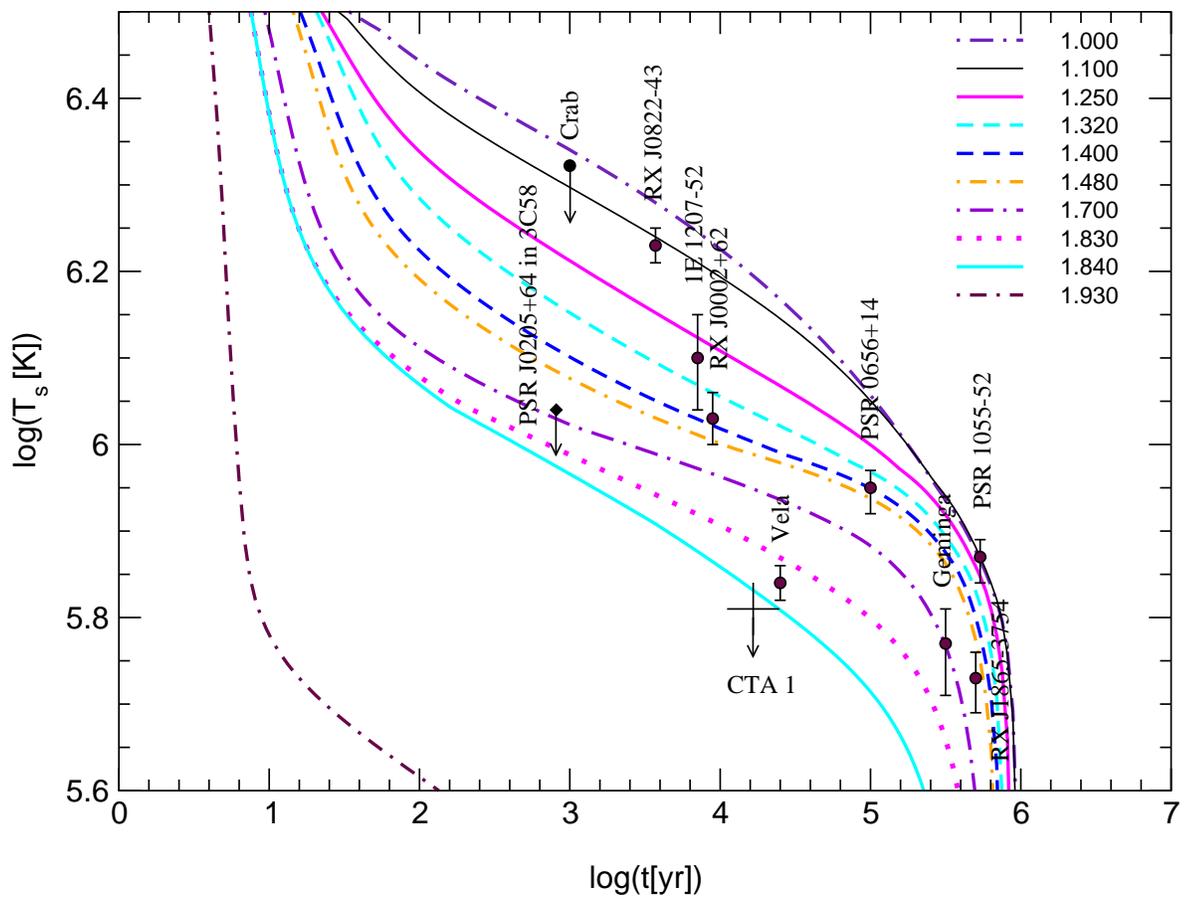,height=0.9\textwidth,angle=-90}}
\caption{Model II. Same as 
%%Fig. \ref{fig:cool-h} but 
%using the Tsuruta law.
%%with the hadronic gaps taken from \cite{TT}, see 
Fig. 20 of Ref. 
\cite{bgv2004}.}
\label{fig:cool-h-tt}
\end{figure}

In \cite{bgv2004} different hadronic cooling models have been presented
which can explain the available cooling data by varying the neutron star
mass in some limits. The question arises whether one of these 
alternatives could be preferred. 
In Ref. \cite{bgv2004} we conjectured that the model based on Fig. 20 of
\cite{bgv2004} (Fig. \ref{fig:cool-h-tt} of the given work)
that has used another
choice of the
gaps  demonstrates the best fit since it shows
a more regular dependence on the neutron star masses. We call it model II.
In this model II the $1S_0$
neutron gap is the same as in the model I used above, whereas the proton $1S_0$
gap is calculated  by \cite{TT} (see  thin lines in Fig. 5 of  \cite{bgv2004}). The neutron $3P_2$ gap is taken from
\cite{TT}
but then is additionally suppressed by factor $0.1$ as required by  the
result of \cite{SF}
and in order to appropriately fit the cooling data.
%%it has been suggested that a cooling model where
%%input data are consistent should be preferred. Consequently, a hadronic
%%cooling model where both the equation of state and the pairing gaps are
%%calculated using the same nucleon-nucleon interaction as a microscopic 
%%input has been suggested on the basis of the Argonne $V18 + \delta v + UIX^*$
%%with the gaps calculated by \cite{TT}, see Fig. \ref{fig:cool-h-tt}. 
According to \cite{pgtb2004} this model II succesfully passed
the mentioned Log N - Log S test supporting the
conjecture of \cite{pgtb2004} and an earlier conjecture of \cite{VS86} that
neutron star masses should be essentially different. 

We have performed the same hybrid cooling calculation as described above, now
for the model II.
The results 
are qualitatively the same as those presented in Figs \ref{fig:cool-2sc} --
\ref{fig:cool-csl-x}. The difference is that now the curves related to low
mass objects ($M<M_{\rm crit}^{\rm DU}$)
are pushed downwards  describing the Crab data point.
However, as in model I, both scenarios either that there are unpaired quarks 
and that the X-gaps are density independent, do not allow to appropriately fit
the data. 

\begin{figure}[ht]
\centerline{
\psfig{figure=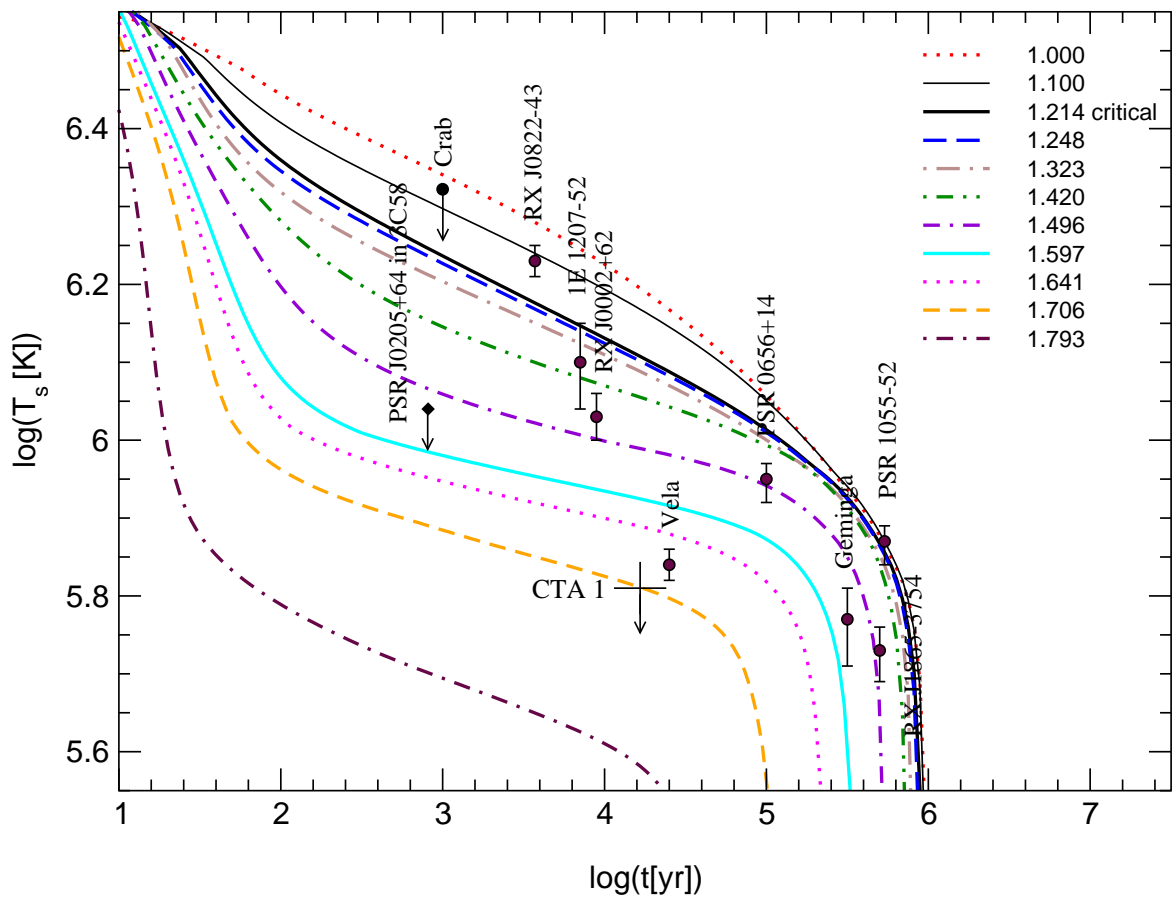,height=0.9\textwidth,angle=-90}}
\caption{Same as Fig. \ref{fig:cool-csl-x} but for model II.
%using the Tsuruta law.
%%with a hadronic cooling model cooresponding to Fig. \ref{fig:cool-h-tt}.
}
\label{fig:cool-csl-x-tt}
\end{figure}

The results for the density-dependent X-gap according to Eq. (\ref{gap}) now 
using the hadronic cooling model II (demonstrated by Fig. \ref{fig:cool-h-tt}) 
are shown in Fig. \ref{fig:cool-csl-x-tt}. 
In comparison with the result of Fig. \ref{fig:cool-csl-x} the variation of 
the cooling data between slow and 
fast coolers is obtained by choosing hybrid star masses in a wider range
$1.1\le M/M_\odot \le 1.7$. 
%%Note that according to a recently suggested independent test of cooling 
%%models \cite{pgtb2004} by comparing results of a corresponding population 
%%synthesis model
%%with the Log N - Log S distribution of nearby isolated X-ray sources
%%the cooling model of Fig. \ref{fig:cool-h-tt} should be preferred 
%over that of Fig. \ref{fig:cool-h} 
%%and thus the results of Fig. \ref{fig:cool-csl-x-tt} for the hybrid stars. 
%%A Log N - Log S test for hybrid star cooling has not yet been paerformed.
Next, it would be interesting to check whether this model passes the 
Log N - Log S test.

\section{Conclusion}
Concluding, we demonstrated that within a recently developed nonlocal, 
chiral quark model
the critical densities for a phase transition to color superconducting
quark matter  can be low enough for these
phases to occur in compact star configurations with masses below
$1.3~M_\odot$.
For the choice of the Gaussian formfactor 
the 2SC - normal quark matter mixed phase arises at $M\simeq 1.21~M_\odot$. 
We have shown that without a residual pairing the 2SC quark matter
phase could describe the cooling data only if  compact stars had
masses in a very narrow band around the critical mass for which the 
quark core can occur. Since there are observations of neutron stars
with higher and essentially different masses such a scenario should be 
disfavored.

Then we assumed that formally unpaired quarks can be paired with small gaps
$\Delta_X <1~$MeV  (2SC+X pairing),
which values we
varied in wide limits. With density independent gaps we failed to
appropriately fit
the cooling data.

We showed that the present day cooling data could be still
explained by hybrid stars, however, when assuming a complex  pairing pattern, 
where quarks are partly strongly paired within the 2SC channel, and partly 
weakly paired with gaps $\Delta_X < 1~$MeV, being rapidly decreasing with the 
increase of the density.
Within the given hypothesis and in the framefork of our quark model for the
equation of state the observational data are explained in a broad mass 
interval 
$M=1.32~-~1.75~M_\odot $ for the model I and 
$M=1.1~-~1.7~M_\odot $ for the model II, as it is favored by the 
experimental data. 
%%For hybrid stars the mass interval  $M=1.32~-~1.75~M_\odot $ is
%%obtained
%%for superconducting quark matter with 2SC+X pairing, where the
%%residual pairing gap has to be a decreasing function of the density.
However we will point out that the result is model dependent and demonstrates
only a
possibility that the presence of the 2SC+X phase might not contradict to the
neutron star cooling data.

%%In summary we have found a scenario of the cooling of a hybrid star with a
%%large quark core being in the
%%quark 2SC+X phase that can reasonably describe the cooling data. 
%%in the interior 
%%an alternative to the hadronic scenario for the compact star cooling data, 
%%a purely hadronic one and a 
%%hybrid star cooling one. As for the hybrid star cooling case, 
We have found rather strong constraints for the 
%%EoS in that the quark matter should be superconducting with a small, 
density dependent X-gap pairing.
It remains to be investigated which microscopic pairing pattern could
fulfill the constraints obtained in this work. Another indirect check of the
model could be the Log N - Log S test.
 
%%\label{introsec}
\subsection*{Acknowledgement}

%%Authors thank E.E. Kolomeitsev for the discussions.
The authors acknowledge the hospitality and support of Rostock
University where most of this work has been performed. 
The work of  H.G.  has been supported in part by the
Virtual Institute of the Helmholtz Association under grant No.
VH-VI-041 and by DAAD University partnership program; 
that of D.V. has been supported in part by DFG grant No. 436 RUS 17/117/03. 

%\begin{thebibliography}{99}

\end{document}